\documentclass[11pt]{article}
\usepackage[margin=1in]{geometry}
\usepackage[T1]{fontenc}
\usepackage{lmodern}
\usepackage{amsmath,amssymb}
\usepackage{booktabs,multirow,tabularx}
\usepackage{graphicx}
\usepackage[numbers,sort&compress]{natbib}
\usepackage{xcolor}
\usepackage[colorlinks=true,linkcolor=blue!50!black,citecolor=blue!50!black,urlcolor=blue!50!black]{hyperref}
\usepackage{caption}
\title{Observability, Not Placement: A PyCPS-IDS Simulation Study of\\ Edge, Fog, Cloud and Hybrid Intrusion Detection in Smart-City IoT}
\author{Mustafa S.\ Aljumaily\thanks{Corresponding author: \href{mailto:mustafa.s@daw-alfada.com}{mustafa.s@daw-alfada.com}}\\ {\small R\&D Department, Daw Alfada Company, Iraq}
\and Hayder Kareem Abd\\ {\small Daw Alfada Company, Iraq}
\and Nawar S.\ Alseelawi\\ {\small University of Misan, Amarah, Iraq}}
\date{}

\begin{document}
\maketitle

\begin{abstract}
Where should intrusion detection run in a smart-city IoT network: on the device, at the district gateway (fog), in the cloud, or across tiers? Recorded datasets cannot answer this, because they fix what each tier observes. PyCPS-IDS is an open-source Python discrete-event simulator that couples traffic, lighting and energy processes with lossy links, finite gateway queues and a lossy backhaul, injects flooding, spoofing, replay and false data injection (FDI) under an explicit threat model, and records what each tier sees. On ten fresh seeds per scenario, evaluated after the pipeline was frozen, fog, cloud and hybrid detectors using the same model reached F1 0.54--0.57, and hybrid and fog were statistically equivalent within $\pm$0.02 in all three city scenarios. Placement changed cost, not detection: the cloud used 827 times the backhaul traffic of fog alerts, its latency followed its batch interval (33\,s median at 30\,s batches, 12\,s streaming, 11\,s at the fog) and it degraded under backhaul loss, while the hybrid cut false alarms on attack-free runs from 54 to 6 per hour. Detection was limited by observability. Flooding was detected in at least 96\% of malicious windows at every tier. Spoofing by an attacker who jams the victim and continues its sequence numbers left no network evidence and was detected in 18--22\% of windows; FDI in 43--84\% on light and energy sensors and almost never on noisy traffic counters; replay of a device's own past readings in 15\%. These attacks call for authentication, redundant sensing and physical watermarking, not a different detector placement.
\end{abstract}


\section{Introduction}
Smart-city services such as adaptive street lighting, traffic monitoring and energy metering depend on large numbers of low-cost networked sensors, and deployed systems of this kind have been shown to be insecure \citep{ghena2014green,ronen2016extended}. An attacker who controls some of these devices, or who can transmit on their radio links, can manipulate the physical picture that operators and control loops act on. Intrusion detection systems (IDS) for such deployments are usually trained and tested on datasets recorded at a fixed vantage point \citep{zarpelao2017survey,ferrag2022edge,koroniotis2019botiot,alsaedi2020toniot}. Recorded datasets cannot answer a design question that every integrator faces: \emph{where} should detection run --- on the device, at the district gateway (fog), in the cloud, or across tiers --- given that each tier sees different data, at different delays, over links that lose packets.

Answering that question needs a simulator in which the same attacks can be replayed while the observation point, the network quality and the sensor noise change, and in which every tier observes only what it would observe in a deployment. This paper presents PyCPS-IDS, an open-source Python simulator built for that comparison, and reports what it shows. Our contributions are:
\begin{enumerate}
  \item A SimPy-based simulator that couples urban physical processes (traffic arrivals, ambient light, adaptive street lighting, energy use) with devices, lossy wireless links, finite gateway queues and a lossy backhaul, and that records separately what the device, the gateway and the cloud each observe (Section~\ref{sec:sim}).
  \item An explicit threat model and four attack classes --- flooding (DoS), identity spoofing on the victim's own link, replay of the device's own past readings, and false data injection (FDI) --- each with three intensity levels and ground-truth labels (Sections~\ref{sec:threat} and \ref{sec:attacks}).
  \item An evaluation protocol with a frozen pipeline, confirmatory test seeds drawn only after freezing, thresholds fixed on validation seeds, bootstrap intervals, Holm-corrected paired tests and equivalence tests, false alarms per hour on attack-free runs, and measured detection latency and backhaul overhead (Section~\ref{sec:protocol}).
  \item A comparison of five detectors (rules, edge, fog, cloud, hybrid) across three city scenarios, three network-degradation levels and two sensor-noise levels, with ablations that separate the effect of each tier's observations from the effect of the model family (Section~\ref{sec:results}).
\end{enumerate}
Every number in this paper is read from result files written by the released code, and every table is generated from those files.

\section{Related Work}\label{sec:related}
\paragraph{IoT and smart-city intrusion detection.} Surveys of IoT intrusion detection classify approaches by detection method and by placement, and note that most are evaluated on recorded traffic \citep{zarpelao2017survey}. SVELTE places lightweight modules on constrained nodes and the main analysis at the border router of a 6LoWPAN network \citep{raza2013svelte}, an early instance of the edge/gateway split studied here. Fog computing was proposed to move computation close to devices for latency and bandwidth reasons \citep{bonomi2012fog}, and distributed fog-level detection has been evaluated for IoT attacks \citep{diro2018distributed}; lightweight online detectors run on gateway-class hardware \citep{mirsky2018kitsune}. Public datasets such as N-BaIoT \citep{meidan2018nbaiot}, Bot-IoT \citep{koroniotis2019botiot}, TON\_IoT \citep{alsaedi2020toniot} and Edge-IIoTset \citep{ferrag2022edge} provide realistic traffic from fixed testbeds, but each fixes where observations were taken and the network conditions they were taken under. Flooding by compromised IoT devices is a documented threat \citep{antonakakis2017mirai}. Identity spoofing in wireless sensor networks can be detected from received signal strength when cryptographic authentication is absent \citep{chen2010identity}; our simulator has no radio channel, so it can test only packet-level spoofing evidence. Our earlier work combined network and physical features for smart-home and smart-city IDS, evaluated on public datasets and a simple simulation \citep{aljumaily2025hybrid}; here the simulator is the experimental instrument.

\paragraph{Physics-based detection and active defenses in CPS.} Attacks that alter sensor values while leaving traffic patterns intact --- false data injection \citep{liu2009fdi} and replay \citep{mo2009replay} --- are invisible to purely network-level monitoring. Physics-based detectors compare measurements with a model of the process or with redundant measurements \citep{giraldo2018survey}, and stealthy attackers can keep their manipulation below such detectors' thresholds \citep{urbina2016stealthy}. Replay of recorded data cannot be detected by passive residual checks when the replayed data are statistically indistinguishable from current data; physical watermarking, in which the controller injects a secret excitation and checks that the sensors reflect it, was proposed for exactly this case \citep{mo2009replay,mo2015watermark}, and dynamic watermarking extends it to networked control systems \citep{satchidanandan2017dynamic}.

\paragraph{Evaluation practice.} \citet{sommer2010outside} argue that anomaly-based network IDS face structural obstacles, including the high cost of false alarms and the difficulty of evaluation; \citet{arp2022dos} catalogue pitfalls in machine-learning security research, including data snooping, inappropriate baselines and thresholds tuned on test data. We follow their recommendations: test seeds are never used for fitting or thresholds, precision and false alarms per hour are reported alongside recall, and intervals are computed across independent simulation seeds.

\paragraph{Simulation frameworks.} ns-3 \citep{riley2010ns3} and OMNeT++ \citep{varga2008omnet} provide packet-level fidelity; Cooja simulates Contiki sensor networks down to the instruction level \citep{osterlind2006cooja}; iFogSim models placement and resource management of applications across IoT, fog and cloud \citep{gupta2017ifogsim} but not attacks or detection; SUMO \citep{lopez2018sumo} provides microscopic traffic; mosaik \citep{schutte2011mosaik} composes smart-grid simulators; physical testbeds such as SWaT \citep{mathur2016swat} provide real process data for one plant. SimPy \citep{simpy} offers process-based discrete-event simulation in plain Python and has been used for smart-city IoT testbeds \citep{karanjkar2019simpy}. PyCPS-IDS trades protocol and radio fidelity for a single Python process in which simulation, feature extraction and model training share data structures, which makes large seed sweeps cheap.

\section{The PyCPS-IDS Simulator}\label{sec:sim}
\begin{table}[t]\centering\small\setlength{\tabcolsep}{5pt}
\caption{Simulation scenarios, network levels and sensor noise levels (values from \texttt{pycps\_ids/config.py}). Runs last 90 simulated minutes; features use 10\,s windows. The moderate network level is used unless stated otherwise.}\label{tab:design}
\begin{tabular}{lrrr}\toprule
Scenario & Intersections & Devices & Gateways\\\midrule
Small district & 4 & 40 & 1 \\
Critical corridor & 12 & 120 & 3 \\
Medium urban area & 16 & 200 & 4 \\
\bottomrule\end{tabular}\\[6pt]
\begin{tabular}{lrrrrr}\toprule
Network level & Link delay (ms) & Link loss (\%) & Backhaul delay (ms) & Backhaul loss (\%) & Outages/h\\\midrule
Mild & 20--40 & 1 & 30--60 & 0.5 & 0 \\
Moderate & 40--80 & 4 & 60--120 & 2 & 1 \\
Severe & 80--120 & 8 & 100--200 & 8 & 4 \\
\bottomrule\end{tabular}\\[6pt]
\begin{tabular}{lrrrr}\toprule
Noise level & Reading noise (\%) & Dropouts (\%) & Stuck-at faults/device-h & Stuck for (s)\\\midrule
Clean & 3--3 & 0 & 0 & 0--0 \\
Messy & 3--7 & 2 & 0.5 & 5--15 \\
\bottomrule\end{tabular}
\end{table}

\subsection{City topology}
A scenario fixes the number of intersections, devices and district gateways (Table~\ref{tab:design}). Intersections are assigned to districts in contiguous blocks, and devices are assigned round-robin to intersections. Each intersection hosts traffic sensors, light sensors and energy meters in a fixed pattern (about 50\%, 30\% and 20\% of devices), so every device has at least one neighbor of the same type at the same intersection. Device parameters (reporting period of 1, 2 or 3\,s; lane share for traffic sensors; calibration gain) are fixed per scenario, as they would be after commissioning; the physical processes, links, noise and attacks vary with the run seed.

\subsection{Physical processes}
Physical processes are computed on a 1\,s grid. Vehicle arrivals at each intersection follow a non-homogeneous Poisson process whose rate combines a per-intersection peak rate (0.3--1.8 vehicles/s), a two-peak daily profile and random micro-bursts (on average six per hour, 1.5--2.5$\times$ for 10--30\,s). Each run starts at a random time of day. A traffic sensor reports the vehicle flow on its lane share (binomial thinning of the intersection arrivals) as a 30\,s moving count in vehicles per minute. Ambient light per district is daylight modulated by a mean-reverting cloud factor; street lighting at each intersection dims with ambient light and brightens with traffic, and brownouts (about one per district per hour, 30--120\,s) reduce lamp output to 70\%. A light sensor measures ambient plus lamp light; an energy meter measures the intersection load, which depends on lamp level and a slowly varying load factor. Readings are therefore correlated within an intersection and, for light and energy, across types.

\subsection{Devices, links and gateways}
Each device is a SimPy process that wakes once per reporting period, reads its sensor with multiplicative Gaussian noise, and transmits a packet (device ID, sequence number, timestamp, value, 40--60 bytes) over its link. Links lose packets independently and delay them by a lognormal amount around a per-link mean. Each district gateway is a single-server queue with finite capacity (25--100 packets) and a service rate of twice the district's nominal offered load; packets arriving at a full queue are dropped. The gateway logs every served packet with the ingress link it arrived on, then forwards it over a lossy backhaul (with optional outages) to the cloud. Table~\ref{tab:design} lists the three network levels and the two noise levels; the \emph{messy} level raises reading noise to 3--7\% and adds 2\% reading dropouts and stuck-at faults (0.5 per device-hour, 5--15\,s).

\subsection{Threat model}\label{sec:threat}
The assets are the sensor readings that drive lighting and traffic decisions, and the availability of the district gateway. We consider two adversaries. \textbf{(A1) Compromised firmware} on some devices can make a device flood the gateway (DoS), scale its readings (FDI) or substitute its own earlier readings (replay). Because the firmware produces the packets, it also produces valid sequence numbers, timestamps and, if devices sign messages, valid signatures; message-level freshness checks therefore do not stop A1 replay. \textbf{(A2) A radio-proximate attacker} without the victim's keys (or on a network without per-packet authentication, common in legacy deployments \citep{ghena2014green}) jams a victim device so that its packets do not reach the gateway, and transmits in its name on the victim's own link, continuing its sequence numbers from overheard traffic (spoofing). The gateway and cloud are trusted. The \textbf{edge monitor} is assumed to run outside the compromised application --- for example in a separate trusted execution environment or on the network interface --- and to see the packets the device actually transmits. If A1 also controls the edge monitor, edge results are an upper bound. Adaptive attackers that shape manipulation to a known detector \citep{urbina2016stealthy}, gateway compromise and routing attacks are out of scope.

\subsection{Attacks}\label{sec:attacks}
Each district receives eight attack episodes per run, rotating through the four types, starting after minute 25 and never overlapping within a district. Intensity (low, medium, high) is drawn per episode. Twenty percent of FDI episodes are followed by a replay on the same devices (a two-stage attack).
\begin{itemize}
  \item \textbf{DoS (flooding, A1):} 5--15 compromised devices transmit at 2--3$\times$, 3--5$\times$ or 5--8$\times$ their normal rate for 30--180\,s, with normal-looking payloads. The gateway queue overflows and drops packets from benign devices too.
  \item \textbf{Spoofing (A2):} for 45--120\,s the attacker jams 1--3 victims and transmits on each victim's link at its reporting period, with sequence numbers continuing the victim's. Low intensity sends a copy of the victim's last overheard reading (a 5\,s-old value with normal noise); medium sends that value biased by +15--25\%; high sends values drawn uniformly from the victim's 5th--95th percentile range. The victim keeps sampling, so its own transmitted stream is unchanged.
  \item \textbf{Replay (A1):} firmware on 10--20 compromised devices replaces current readings with the device's own readings from 5--10 (low), 10--20 (medium) or 20--25 (high) minutes earlier for 60--200\,s, leaving packet timing unchanged.
  \item \textbf{FDI (A1):} firmware on 5--10 compromised devices scales readings by $\pm$5--10\% (low), $\pm$15--25\% ramped over half the episode (medium) or $\pm$40--60\% abruptly (high) for 20--90\,s.
\end{itemize}
A device-window is labeled malicious when an attack targeting that device overlaps it by at least half the window. For DoS, the flooding devices are the positives; benign devices that lose packets in the congested queue are negatives, so false alarms on collateral damage are counted.

\section{Detectors and Evaluation Protocol}\label{sec:protocol}
\subsection{What each tier observes}
Features are computed per device over 10\,s windows (Table~\ref{tab:features}), and each tier computes features only from its own log. The \textbf{edge} sees the device's own transmitted stream: send rate, and the level, jump and variability of its own values against its last 2 minutes. The \textbf{fog} (district gateway) sees received packets: count relative to the expected rate, duplicate and out-of-order sequence numbers, sequence gaps, number of distinct ingress links and share of packets on a link other than the registered one, inter-arrival variability, queue length and drops; and physical features: each device's deviation from its own 2-minute history and from the median of same-type devices at its intersection, expressed as a robust z-score against the ratio and spread learned on attack-free training windows. The \textbf{cloud} sees the packets that survived the backhaul; it keeps longer history (10-minute self history, 60\,s aggregates of neighbor residual, duplicates and link mismatch), a missing-data flag, and one feature no gateway can compute: each device's deviation from the city-wide median of its sensor type, which draws on all districts. It cannot see queue length or drops. The \textbf{hybrid} decides at the fog with fog features, the edge model's score, and the cloud's long-horizon context for the previous window, computed from 20-byte per-window fog summaries; a summary lost on the backhaul leaves the last received context in place and sets a missing flag. All three tiers also receive the device's sensor type.

\begin{table}[t]\centering\small
\caption{Features per device and 10\,s window, by the tier that can compute them.}\label{tab:features}
\begin{tabularx}{\linewidth}{lX}\toprule
Tier & Features\\\midrule
Edge & own send rate / expected rate; level of own values vs.\ own last 2\,min ($z$-score); jump from previous window; variability ratio\\
Fog (network) & received count / expected; duplicate sequence numbers; out-of-order packets; sequence gaps; distinct ingress links; share of packets on an unregistered link; inter-arrival coefficient of variation; gateway queue length; queue drops\\
Fog (physical) & $z$-score vs.\ own last 2\,min; jump; variability ratio; log-ratio to the median of same-type neighbors at the intersection, as a robust $z$-score against the median and MAD learned on attack-free training windows (neighbor residual)\\
Cloud & fog network and physical features recomputed from packets that survived the backhaul (no queue length or drops); 10\,min self history; 60\,s means of neighbor residual, duplicates and link mismatch; missing-data flag; deviation from the city-wide median of the same sensor type (all districts)\\
Hybrid & fog features + edge model score + cloud context for the previous window (10\,min $z$-score, 60\,s neighbor residual, duplicates and link mismatch, missing flag), computed from fog summaries\\
All tiers & sensor type (traffic, light, energy), from the device registry\\
\bottomrule\end{tabularx}
\end{table}

\subsection{Models}
The \textbf{rules} baseline normalizes each of seven fog features by its 99.9th percentile over attack-free training windows and scores a window by the largest normalized value; like every detector, its alert threshold on that score is chosen on the validation seeds. The \textbf{edge} model is a depth-4 decision tree, sized for a microcontroller. The \textbf{fog}, \textbf{cloud} and \textbf{hybrid} models are the same random forest \citep{breiman2001rf} (100 trees, maximum depth 14, minimum leaf size 3, balanced class weights, half-sample bootstrap), so that differences between them come from what each tier observes and not from the model family. An ablation replaces the cloud forest with a multilayer perceptron (64 and 32 hidden units) trained on oversampled positives. The hybrid is trained on \emph{cross-fitted} edge scores: the training seeds are split in two halves, an edge model is trained on each half, and each half's edge scores come from the model trained on the other half, so the hybrid never learns from edge scores that have seen their own labels. Further ablations train the fog model on network features only and on physical features only, and the hybrid without the edge score or without the cloud context. All models use scikit-learn \citep{pedregosa2011sklearn}.

\subsection{Protocol}\label{sec:proto}
For each scenario, models are trained on seeds 1--4, and each model's decision threshold is set to maximize F1 on the pooled validation seeds 5--6; thresholds are then fixed. Each run simulates 90 minutes; the first 5 minutes only feed history, and attacks start after minute 25. Per-device neighbor and city references are learned on attack-free windows of the training seeds.

\paragraph{Development and confirmatory seeds.} The simulator, features and models were developed over several iterations using seeds 101--110 for evaluation; design changes were made after seeing those results (Section~\ref{sec:history}). The pipeline was then frozen, and all results in this paper come from ten \emph{confirmatory} test seeds (201--210) per scenario that were simulated only after freezing. (A 40-minute smoke test of the small scenario touched two of these seeds to check that the scripts ran; its results were not used, and no code changed after it.)

\paragraph{Metrics and statistics.} We report true-positive rate (TPR), false-positive rate (FPR), precision, F1, macro-F1 over the four attack types, ROC-AUC and PR-AUC per test seed, and give means with 95\% percentile-bootstrap intervals over the ten seeds. Per-attack results use the positives of one attack type together with all negatives. Detectors are compared by paired differences on the same seeds; we report two-sided paired $t$-tests with Holm correction over the whole family of 39 comparisons, and paired TOST equivalence tests with a margin of $\pm$0.02, also Holm-corrected. Two detectors are called \emph{equivalent} only when the TOST test rejects, not merely when a difference test fails to. False alarms per hour are counted on three additional attack-free runs of the medium area.

\textbf{Latency} is measured per attacked device: the time from attack start until the tier's first alert on a positive window, where the edge decides at the end of the window, the fog after the device-to-gateway delay, and the cloud after the gateway-to-cloud delay at its next batch (every 30\,s by default; 0 and 5\,s are also reported). \textbf{Overhead} is the IDS-related traffic on the backhaul: alert messages (32 bytes) for edge, fog and rules; every forwarded packet plus a 20-byte header for the cloud; summaries plus alerts for the hybrid. \textbf{Compute cost} is the per-window inference time on one CPU core and the serialized model size (not run-time memory).

\section{Results}\label{sec:results}
\subsection{Overall detection}
Table~\ref{tab:main} gives detection on the confirmatory test seeds. In all three scenarios the fog, cloud and hybrid detectors reached similar F1 (0.54--0.57); the largest mean difference between any of them and the fog was 0.018. Table~\ref{tab:stats} gives the paired comparisons. The hybrid was equivalent to the fog within $\pm$0.02 F1 in 3 of three scenarios. The cloud was equivalent to the fog in 1 (medium); in the other scenarios its interval was too wide to show equivalence, and no difference survived Holm correction. Small differences can be statistically detectable without being practically relevant: in the corridor the hybrid's +0.010 F1 over the fog was significant after correction ($p$=0.008) and also inside the equivalence margin. The hybrid ranked windows better than the fog in every scenario (ROC-AUC +0.021--+0.026; Table~\ref{tab:stats}), but at the operating threshold this did not translate into higher F1.

Fog improved on the percentile-rule baseline by 0.054--0.065 F1 and on its own network-only variant by 0.083--0.091. The edge detector, which sees only the device's own stream, reached F1 0.45--0.47 with precision 0.96 and the lowest recall. Recall is limited by attack type rather than placement: replay, FDI and spoofing make up 70\% of malicious windows in the medium area, and Section~\ref{sec:attackres} shows that most of them leave no evidence that any tier can see.

\begin{table}[t]\centering\small
\caption{Detection on held-out test seeds (moderate network, clean sensors). Means over the 10 confirmatory test seeds (201--210) with 95\% bootstrap intervals in brackets. Thresholds were fixed on validation seeds 5--6.}\label{tab:main}
\begin{tabular}{lrrrrrr}\toprule
Detector & TPR & FPR & Precision & F1 & ROC-AUC & PR-AUC\\\midrule
\multicolumn{7}{l}{\textit{Small district} (6,691 malicious of 204,000 test device-windows)}\\
Rules & 0.35 \scriptsize{[0.29, 0.41]} & 0.002 \scriptsize{[0.001, 0.003]} & 0.86 \scriptsize{[0.82, 0.90]} & 0.49 \scriptsize{[0.42, 0.55]} & 0.69 & 0.36\\
Edge & 0.31 \scriptsize{[0.25, 0.37]} & 0.000 \scriptsize{[0.000, 0.001]} & 0.96 \scriptsize{[0.94, 0.98]} & 0.46 \scriptsize{[0.40, 0.53]} & 0.70 & 0.37\\
Fog & 0.43 \scriptsize{[0.38, 0.48]} & 0.004 \scriptsize{[0.003, 0.005]} & 0.78 \scriptsize{[0.75, 0.81]} & 0.55 \scriptsize{[0.51, 0.60]} & 0.77 & 0.48\\
Cloud & 0.43 \scriptsize{[0.38, 0.49]} & 0.002 \scriptsize{[0.002, 0.003]} & 0.86 \scriptsize{[0.83, 0.89]} & 0.57 \scriptsize{[0.52, 0.62]} & 0.79 & 0.50\\
Hybrid & 0.45 \scriptsize{[0.40, 0.50]} & 0.006 \scriptsize{[0.005, 0.007]} & 0.72 \scriptsize{[0.67, 0.76]} & 0.55 \scriptsize{[0.51, 0.59]} & 0.79 & 0.49\\
\midrule
\multicolumn{7}{l}{\textit{Critical corridor} (20,945 malicious of 612,000 test device-windows)}\\
Rules & 0.34 \scriptsize{[0.32, 0.37]} & 0.002 \scriptsize{[0.002, 0.003]} & 0.86 \scriptsize{[0.82, 0.89]} & 0.49 \scriptsize{[0.46, 0.52]} & 0.67 & 0.35\\
Edge & 0.31 \scriptsize{[0.29, 0.35]} & 0.000 \scriptsize{[0.000, 0.001]} & 0.96 \scriptsize{[0.95, 0.97]} & 0.47 \scriptsize{[0.44, 0.51]} & 0.70 & 0.37\\
Fog & 0.43 \scriptsize{[0.40, 0.47]} & 0.005 \scriptsize{[0.005, 0.006]} & 0.73 \scriptsize{[0.71, 0.76]} & 0.54 \scriptsize{[0.51, 0.57]} & 0.79 & 0.49\\
Cloud & 0.43 \scriptsize{[0.40, 0.47]} & 0.005 \scriptsize{[0.004, 0.006]} & 0.75 \scriptsize{[0.73, 0.77]} & 0.55 \scriptsize{[0.52, 0.58]} & 0.80 & 0.50\\
Hybrid & 0.45 \scriptsize{[0.42, 0.48]} & 0.006 \scriptsize{[0.005, 0.006]} & 0.73 \scriptsize{[0.71, 0.75]} & 0.55 \scriptsize{[0.53, 0.58]} & 0.81 & 0.50\\
\midrule
\multicolumn{7}{l}{\textit{Medium urban area} (28,982 malicious of 1,020,000 test device-windows)}\\
Rules & 0.34 \scriptsize{[0.31, 0.38]} & 0.002 \scriptsize{[0.002, 0.002]} & 0.85 \scriptsize{[0.83, 0.87]} & 0.49 \scriptsize{[0.45, 0.52]} & 0.68 & 0.35\\
Edge & 0.30 \scriptsize{[0.27, 0.33]} & 0.000 \scriptsize{[0.000, 0.000]} & 0.96 \scriptsize{[0.95, 0.97]} & 0.45 \scriptsize{[0.42, 0.49]} & 0.68 & 0.34\\
Fog & 0.43 \scriptsize{[0.39, 0.46]} & 0.004 \scriptsize{[0.004, 0.005]} & 0.74 \scriptsize{[0.72, 0.77]} & 0.54 \scriptsize{[0.51, 0.57]} & 0.78 & 0.47\\
Cloud & 0.43 \scriptsize{[0.40, 0.46]} & 0.004 \scriptsize{[0.004, 0.005]} & 0.74 \scriptsize{[0.71, 0.76]} & 0.54 \scriptsize{[0.51, 0.57]} & 0.80 & 0.49\\
Hybrid & 0.43 \scriptsize{[0.40, 0.46]} & 0.004 \scriptsize{[0.004, 0.005]} & 0.75 \scriptsize{[0.72, 0.78]} & 0.55 \scriptsize{[0.52, 0.57]} & 0.81 & 0.49\\
\bottomrule
\end{tabular}
\end{table}

\begin{table}[t]\centering\small
\caption{Paired differences on the same 10 test seeds: mean, 95\% bootstrap interval, Holm-adjusted two-sided $p$ (family of 39 tests), and whether a Holm-adjusted paired TOST test shows equivalence within $\pm$0.02. Upper block: F1; lower block: ROC-AUC.}\label{tab:stats}
\begin{tabular}{llrrrc}\toprule
Scenario & Comparison & $\Delta$ & 95\% interval & $p_\text{Holm}$ & Equivalent\\\midrule
Small & hybrid $-$ fog & -0.003 & [-0.009, +0.003] & 1.000 & yes\\
Small & cloud $-$ fog & +0.018 & [+0.001, +0.039] & 1.000 & no\\
Small & cloud (MLP) $-$ cloud & -0.011 & [-0.020, -0.002] & 0.519 & no\\
Small & fog $-$ fog (network only) & +0.089 & [+0.063, +0.120] & 0.008 & no\\
Small & fog $-$ rules & +0.065 & [+0.038, +0.099] & 0.071 & no\\
Corridor & hybrid $-$ fog & +0.010 & [+0.007, +0.013] & 0.008 & yes\\
Corridor & cloud $-$ fog & +0.005 & [-0.007, +0.018] & 1.000 & no\\
Corridor & cloud (MLP) $-$ cloud & -0.019 & [-0.029, -0.009] & 0.121 & no\\
Corridor & fog $-$ fog (network only) & +0.083 & [+0.071, +0.094] & $<$0.001 & no\\
Corridor & fog $-$ rules & +0.055 & [+0.046, +0.063] & $<$0.001 & no\\
Medium & hybrid $-$ fog & +0.007 & [+0.003, +0.010] & 0.135 & yes\\
Medium & cloud $-$ fog & +0.000 & [-0.006, +0.006] & 1.000 & yes\\
Medium & cloud (MLP) $-$ cloud & -0.011 & [-0.026, +0.002] & 1.000 & no\\
Medium & fog $-$ fog (network only) & +0.091 & [+0.074, +0.109] & $<$0.001 & no\\
Medium & fog $-$ rules & +0.054 & [+0.039, +0.069] & 0.003 & no\\
\midrule
Small & hybrid $-$ fog & +0.023 & [+0.015, +0.031] & 0.016 & no\\
Small & cloud $-$ fog & +0.022 & [+0.010, +0.035] & 0.135 & no\\
Corridor & hybrid $-$ fog & +0.021 & [+0.018, +0.025] & $<$0.001 & no\\
Corridor & cloud $-$ fog & +0.015 & [+0.008, +0.022] & 0.070 & no\\
Medium & hybrid $-$ fog & +0.026 & [+0.022, +0.031] & $<$0.001 & no\\
Medium & cloud $-$ fog & +0.021 & [+0.015, +0.026] & 0.002 & no\\
\bottomrule\end{tabular}
\end{table}

\subsection{Detection by attack type}\label{sec:attackres}
\begin{figure}[t]\centering
\includegraphics[width=\linewidth]{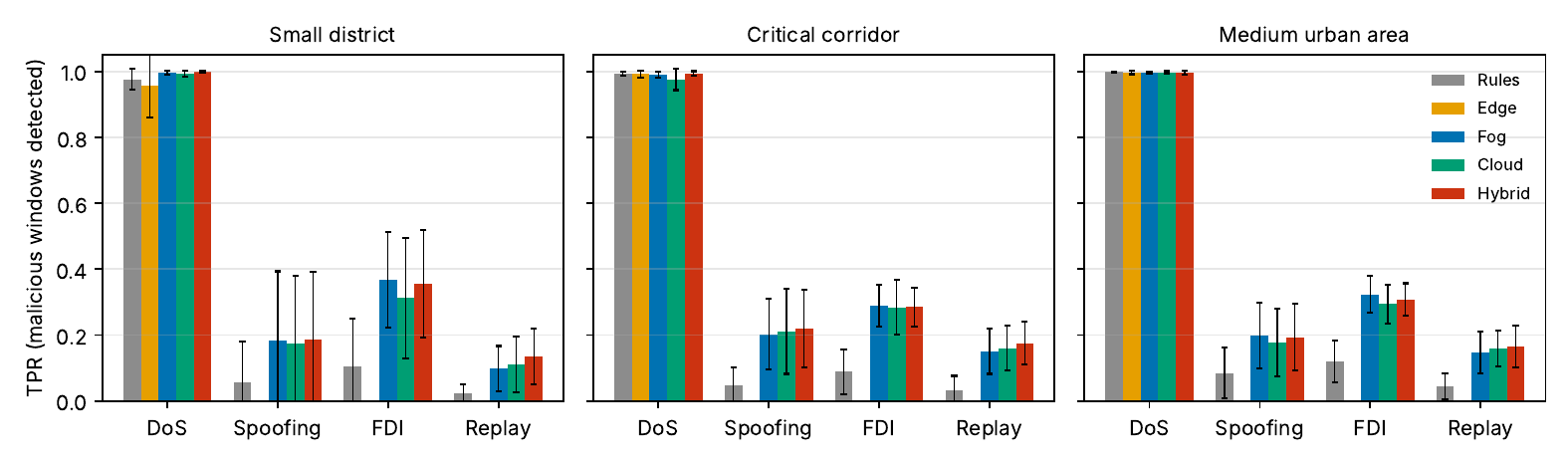}
\caption{Share of malicious windows detected by attack type and detector, per scenario (mean over 10 test seeds; error bars are one standard deviation across seeds).}\label{fig:attack}
\end{figure}

Figure~\ref{fig:attack} and Table~\ref{tab:attack} break detection down by attack. \textbf{Flooding} was detected in 96\% or more of malicious windows by every tier, including the edge, because a flooding device sees its own send rate rise. \textbf{Spoofing} was detected in 18--22\% of windows at the fog and above and never at the edge. Because the attacker jams the victim, transmits on the victim's link and continues its sequence numbers, the gateway sees one well-formed stream, and the network-only fog variant detected 0--3\% of spoofed windows. What remains is the content of the injected values. Table~\ref{tab:breakdown} shows that spoofing with biased values (medium intensity) was caught in 88--100\% of windows on light and energy sensors and spoofing with random in-range values (high) in 50--55\%, while copying the victim's last overheard reading (low) was never caught. \textbf{FDI} was detected in 32\% and \textbf{replay} in 15\% of malicious windows by the fog in the medium area.

The network-only fog ablation detected flooding as well as the full detector and almost no FDI, replay or spoofing; the physical-only ablation did the reverse (Table~\ref{tab:attack}).

\begin{table}[t]\centering\small
\caption{Share of malicious device-windows detected (TPR) by attack type, medium urban area, mean over 10 test seeds. The last two rows are fog ablations.}\label{tab:attack}
\begin{tabular}{lrrrr}\toprule
Detector & DoS & Spoofing & FDI & Replay\\\midrule
Rules & 1.00 & 0.08 & 0.12 & 0.04\\
Edge & 1.00 & 0.00 & 0.00 & 0.00\\
Fog & 1.00 & 0.20 & 0.32 & 0.15\\
Cloud & 1.00 & 0.18 & 0.29 & 0.16\\
Hybrid & 1.00 & 0.19 & 0.31 & 0.16\\
Fog, network only & 1.00 & 0.03 & 0.00 & 0.00\\
Fog, physical only & 0.00 & 0.20 & 0.31 & 0.14\\\midrule
Malicious windows & 8,744 & 1,301 & 3,549 & 15,388\\
\bottomrule\end{tabular}
\end{table}

\paragraph{Which value attacks are detectable.} Table~\ref{tab:breakdown} splits the physical-only fog detector's recall by sensor type and intensity. On light sensors and energy meters, FDI was detected in 43--84\% of windows, rising with intensity, and replay in 20--37\%. On traffic counters, value attacks were detected in at most 2\% of windows at any intensity: a counter reports a 30\,s count of a few to a few dozen vehicles, so its benign window-to-window variation exceeds even a 40--60\% bias, and at the chosen threshold no per-window detector separated the two. Traffic counters are half of all devices, so this one sensor type bounds the achievable recall for value attacks.

\begin{table}[t]\centering\small
\caption{Recall of the physical-only fog detector by attack, intensity and sensor type (medium urban area, 10 test seeds). Spoofing intensities: low = copy of the victim's last overheard reading; medium = that reading biased by +15--25\%; high = random values in the victim's normal range.}\label{tab:breakdown}
\begin{tabular}{llrrr}\toprule
Attack & Intensity & Light & Energy & Traffic\\\midrule
FDI & low & 0.50 & 0.43 & 0.00\\
FDI & medium & 0.76 & 0.74 & 0.00\\
FDI & high & 0.84 & 0.82 & 0.01\\
Replay & low & 0.35 & 0.20 & 0.00\\
Replay & medium & 0.34 & 0.37 & 0.00\\
Replay & high & 0.29 & 0.37 & 0.00\\
Spoofing & low & 0.00 & 0.00 & 0.01\\
Spoofing & medium & 0.88 & 1.00 & 0.02\\
Spoofing & high & 0.50 & 0.55 & 0.00\\
\bottomrule\end{tabular}
\end{table}

\subsection{Latency, overhead, false alarms and compute}
\begin{figure}[t]\centering
\begin{minipage}{0.49\linewidth}\centering\includegraphics[width=\linewidth]{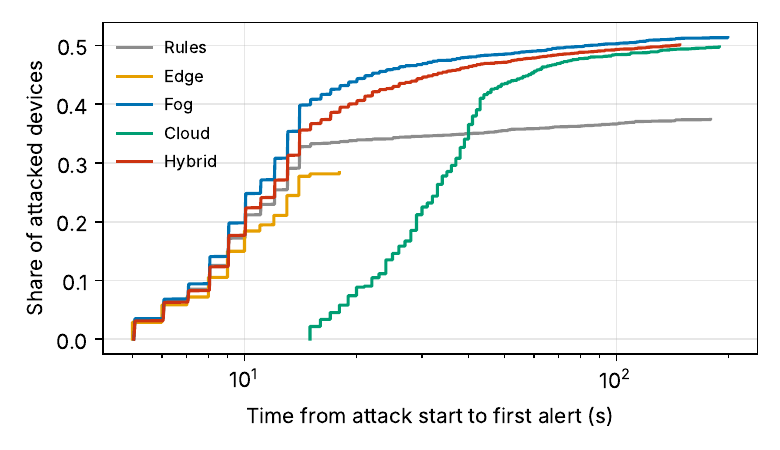}\caption{Time from attack start to first alert, medium area (cloud: 30\,s batch). Curves end at the share of attacked devices detected.}\label{fig:latency}\end{minipage}\hfill
\begin{minipage}{0.49\linewidth}\centering\includegraphics[width=\linewidth]{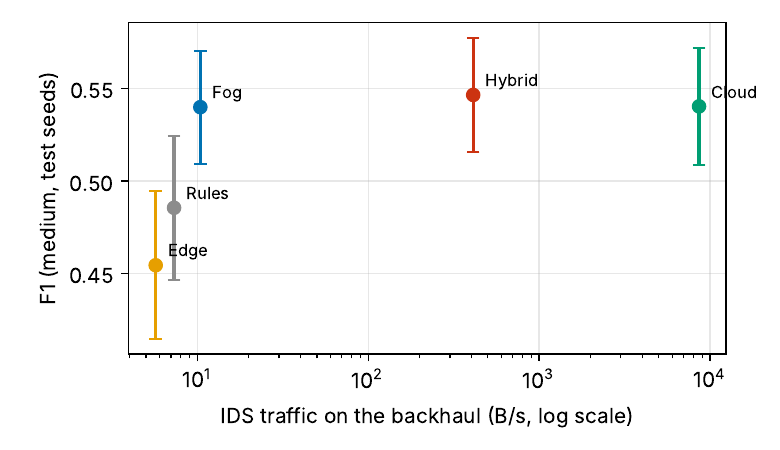}\caption{F1 against IDS traffic on the backhaul, medium area (95\% intervals over seeds).}\label{fig:overhead}\end{minipage}
\end{figure}

Table~\ref{tab:ops} reports operational costs for the medium area. Fog, cloud and hybrid alerted on about the same share of attacked devices (0.50--0.51). The cloud's median time to first alert was 33\,s with its default 30\,s batch, 17\,s with 5\,s batches and 12\,s when streaming, against 11\,s for the fog (Figure~\ref{fig:latency}): the cloud's delay comes from how it batches, not from the backhaul. For flooding the fog alerted within one window (median 9\,s); for the FDI attacks it detected, the median was 16\,s. Streaming every reading to the cloud generated 8,605\,B/s of traffic on the backhaul that exists only for detection, against 10\,B/s of fog alerts and 411\,B/s for the hybrid's summaries and alerts --- a 21-fold difference between cloud and hybrid at equal detection (Figure~\ref{fig:overhead}). If the readings are sent to the cloud anyway for other purposes, the cloud's incremental IDS traffic is close to zero; the comparison then reduces to latency, robustness and false alarms.

False alarms separated the detectors more than detection did. On attack-free runs of the medium area the fog raised 53.6 false alarms per hour for the whole city, the rule baseline 28.9, the cloud 9.2, the hybrid 6.1 and the edge 0.0. During attack runs the false alarm rate was higher for every detector (about 301 per hour for the fog), because most false positives fall on benign devices next to an attack (for example, devices whose packets are dropped by a flooded queue). The forest-based fog, cloud and hybrid models took 3.8--4.8\,$\mu$s per device-window and serialized to 11--12\,MB; the edge tree took 0.04\,$\mu$s and 3\,kB.

\begin{table}[t]\centering\small
\caption{Operational cost, medium urban area (10 test seeds). \emph{Attacked devices detected}: share of (episode, target) pairs with at least one alert during the attack. Latency is from attack start to the first alert, over detected pairs. Backhaul overhead is IDS-related traffic between gateways and cloud for the whole city. \emph{False alarms/h}: alerts per hour for the whole city on three attack-free runs. The cloud row uses a 30\,s batch; the indented rows repeat its latency for 5\,s batches and streaming. Inference time is per device-window on one CPU core (batch prediction); model size is the serialized size, not run-time memory.}\label{tab:ops}
\footnotesize\setlength{\tabcolsep}{4pt}
\begin{tabular}{lrrrrrrr}\toprule
 & Attacked devices & \multicolumn{2}{c}{Latency (s)} & Backhaul & False & Inference & Model\\
Detector & detected & median & 90th pct. & (B/s) & alarms/h & ($\mu$s/window) & (kB)\\\midrule
Rules & 37\% & 10.1 & 18.3 & 7.3 & 28.9 & 0.03 & 0\\
Edge & 28\% & 9.0 & 14.0 & 5.7 & 0.0 & 0.04 & 3\\
Fog & 51\% & 11.1 & 26.1 & 10.4 & 53.6 & 3.78 & 10,781\\
Cloud & 50\% & 33.0 & 57.0 & 8,605.0 & 9.2 & 4.25 & 12,229\\
\quad batch 5\,s & 50\% & 17.0 & 38.0 & & & & \\
\quad batch 0\,s (streaming) & 50\% & 12.1 & 33.1 & & & & \\
Hybrid & 50\% & 12.1 & 33.0 & 410.6 & 6.1 & 4.76 & 11,814\\
\bottomrule\end{tabular}
\end{table}

\subsection{Robustness to network degradation and sensor faults}
With models trained at the moderate network level, the severe level (8\% link loss, 8\% backhaul loss, about four backhaul outages per hour) left the cloud without data for 6\% of device-windows and lowered its F1 from 0.540 to 0.517; fog and hybrid F1 changed by at most 0.002 (Table~\ref{tab:robust}). The hybrid keeps deciding at the fog when cloud context is missing, which is why it stayed at the fog's level.

Sensor faults had a larger effect. Applied to messy runs (more noise, dropouts and stuck-at faults), models trained on clean runs lost 28--35\% of their F1, mainly through more false positives (fog FPR 0.0043 on clean and 0.0227 on messy runs). Re-choosing only the thresholds on messy validation runs, without retraining, restored F1 to 0.48--0.49, and retraining on messy runs to 0.52. The rule baseline (0.49) and the edge tree (0.45) were unaffected.

\begin{table}[t]\centering\small
\caption{F1 in the medium urban area under network degradation (models trained at the moderate level) and under messy sensors: clean-trained models; the same models with thresholds re-chosen on messy validation runs (\emph{recalibrated}); and models retrained on messy runs. Mean over 10 test seeds.}\label{tab:robust}
\setlength{\tabcolsep}{4pt}
\begin{tabular}{lrrrrrr}\toprule
 & \multicolumn{3}{c}{Network level} & \multicolumn{3}{c}{Messy sensors}\\\cmidrule(lr){2-4}\cmidrule(lr){5-7}
Detector & Mild & Moderate & Severe & clean-trained & recalibrated & messy-trained\\\midrule
Rules & 0.49 & 0.49 & 0.49 & 0.49 & 0.49 & 0.49\\
Edge & 0.45 & 0.45 & 0.45 & 0.45 & 0.45 & 0.45\\
Fog & 0.54 & 0.54 & 0.54 & 0.39 & 0.48 & 0.52\\
Cloud & 0.54 & 0.54 & 0.52 & 0.35 & 0.49 & 0.52\\
Hybrid & 0.55 & 0.55 & 0.54 & 0.37 & 0.49 & 0.52\\\midrule
No cloud data & 0.0\% & 0.6\% & 5.9\% & & & \\
\bottomrule\end{tabular}
\end{table}

\subsection{Ablations}
Removing the edge score or the cloud context from the hybrid changed F1 by at most 0.009 in any scenario (Table~\ref{tab:ablation}). Replacing the cloud's random forest by a multilayer perceptron lowered cloud F1 by 0.011--0.019: with a different model family at the cloud, a placement comparison would have attributed part of that gap to placement.

\begin{table}[t]\centering\small
\caption{Ablations: F1 per scenario (mean over 10 test seeds).}\label{tab:ablation}
\begin{tabular}{lrrr}\toprule
Detector & Small & Corridor & Medium\\\midrule
Fog & 0.55 & 0.54 & 0.54\\
Fog, network only & 0.46 & 0.46 & 0.45\\
Fog, physical only & 0.17 & 0.19 & 0.19\\
Cloud & 0.57 & 0.55 & 0.54\\
Cloud, MLP & 0.56 & 0.53 & 0.53\\
Hybrid & 0.55 & 0.55 & 0.55\\
Hybrid, no edge score & 0.55 & 0.55 & 0.54\\
Hybrid, no cloud context & 0.55 & 0.54 & 0.54\\
\bottomrule\end{tabular}
\end{table}

\subsection{Development history}\label{sec:history}
Table~\ref{tab:history} lists the three versions of the pipeline developed for this paper; all of them differ from the earlier preprint \citep{aljumaily2026pycpsrg}, which used a different simulator design and evaluation protocol. In v0, physical features entered the models in raw units and without the sensor type, so noisy traffic counters masked the informative light and energy sensors (physical-only F1 0.08). v1 added per-device scaling learned on training seeds and the sensor type. v2 made the spoofing attacker transmit on the victim's link with jamming and continued sequence numbers (in v0 and v1 it used its own link, which made spoofing nearly trivial to detect), gave the cloud the same model family as the fog and a city-wide feature, cross-fitted the hybrid's edge scores, set replay offsets by intensity and removed unused timestamp jitter. Each change was made after looking at results on development seeds. The ranking of fog, cloud and hybrid changed between versions by amounts of the order of 0.01--0.02 F1, which is why we report only differences that hold on confirmatory seeds and treat smaller ones as ties.

\begin{table}[t]\centering\small
\caption{Development history, medium urban area: F1 of the fog, cloud and hybrid detectors, F1 of the physical-only fog detector, and fog TPR for spoofing and FDI. v0 and v1 were evaluated on development seeds; v2 is the frozen pipeline on confirmatory seeds. Result files of every version are in the repository.}\label{tab:history}
\begin{tabular}{llrrrrrr}\toprule
Version & Seeds & Fog & Cloud & Hybrid & Fog, physical & Spoof TPR & FDI TPR\\\midrule
v0 & 101--110 & 0.472 & 0.477 & 0.489 & 0.08 & 0.99 & 0.01\\
v1 & 101--110 & 0.543 & 0.541 & 0.543 & 0.18 & 0.99 & 0.32\\
v2 (this paper) & 201--210 & 0.540 & 0.540 & 0.547 & 0.19 & 0.20 & 0.32\\
\bottomrule\end{tabular}
\end{table}

\section{Discussion}\label{sec:discussion}
\paragraph{Observability decides detection; placement decides cost.} Each attack class was detected by the tiers that observe its evidence and missed by the others. Flooding changes the attacker's own send rate, so even the edge saw it. FDI and value-changing spoofing are visible only as disagreement between neighboring sensors, which needs at least the fog's district view. With the same model family at every tier, giving a detector more history (cloud), a city-wide view (cloud) or more inputs (hybrid) left F1 within a few hundredths of the fog's. What placement did change was cost: backhaul traffic, latency (driven by the cloud's batching), robustness to backhaul loss, and false alarms. For deployments resembling our scenarios this suggests the gateway as the primary decision point; a hybrid that adds cloud context to fog decisions if the false alarm burden matters (54 against 6 per hour here, for 411\,B/s of summaries); and the cloud for forensics and retraining, streaming if it must alert.

\paragraph{Easy spoofing is an artifact of an easy attacker.} In v1, where the attacker transmitted on its own link, spoofing was detected in 99\% of windows from link mismatch and duplicate sequence numbers. An attacker who jams the victim and continues its sequence numbers removes both signals, and detection fell to 20\%. Spoofing then becomes a data-integrity problem: it is caught when the injected values disagree with neighbors and not when they copy the truth. A copying attacker does no immediate harm, but it holds the device's identity for later manipulation. Reliable detection needs per-packet authentication or radio-level evidence such as signal strength \citep{chen2010identity}, which our simulator does not model. Spoofing results from simulators or testbeds in which the attacker does not suppress the victim should be read as upper bounds.

\paragraph{Physical consistency is redundancy.} FDI detection here is detection of disagreement with redundant neighbors. It worked where neighboring sensors measure nearly the same quantity with little noise (light, energy) and not where they do not (traffic counts). Designers who want FDI detection on such sensors need redundant, lower-noise measurements of the same quantity, or a process model whose residuals are tighter than the bias they want to detect \citep{giraldo2018survey,urbina2016stealthy}.

\paragraph{Replay needs an active defense.} A reading replayed from minutes earlier differs from the truth only by how much the world has changed since, so passive detectors catch replay only when conditions changed quickly. Because compromised firmware (A1) produces fresh sequence numbers, timestamps and signatures, message freshness does not help either. The defense is physical watermarking \citep{mo2015watermark,satchidanandan2017dynamic}: the lighting controller adds a small secret dimming pattern to each lamp, and a light sensor whose readings do not reflect it is reporting something other than the present. Replay by an on-path attacker, which we did not model, is stopped by authenticated freshness. In neither case does detector placement help.

\paragraph{Most of the cost of sensor faults was a threshold problem.} Clean-trained detectors lost much of their F1 on faulty sensors, but re-choosing thresholds on a small set of messy validation runs recovered most of it without retraining. This is a concrete instance of the distribution-shift pitfall described by \citet{arp2022dos}. In operation, it argues for periodically recalibrating thresholds on recent benign data and for training on degraded conditions, which simulation makes cheap.

\section{Limitations}\label{sec:limits}
The physical models are deliberately simple: Poisson traffic, a sinusoidal daylight profile and a linear lighting load. They produce correlated, noisy time series with daily structure, but they are not calibrated to any real city, and the absolute detection rates in this paper should not be read as field performance. Links are modeled by independent loss and lognormal delay, not by a radio channel or a MAC protocol, so radio-level evidence (signal strength, jamming detection) is absent; the spoofing attacker is assumed to jam perfectly and to overhear the victim's readings. Attackers are not adaptive: no attacker shapes its manipulation to stay below a known detector, which \citet{urbina2016stealthy} show is possible, so recall for value attacks is optimistic against such an attacker. The edge monitor is assumed to be outside the compromised firmware. Each scenario is trained and tested on the same topology, with one window length (10\,s) and one labeling rule (at least half-window overlap); we did not sweep them. Model hyperparameters were set once for all tiers and not tuned. Ten seeds per scenario resolve F1 differences of about 0.01--0.02, and false alarms per hour come from three attack-free runs of one scenario. The latency model treats computation time as negligible. All of these are parameters of the released code.

\section{Conclusion}
PyCPS-IDS makes it possible to ask where intrusion detection should run in a smart-city IoT network and to answer with measured detection, latency, overhead and false alarms instead of assumptions. In our scenarios, with the same model at every tier, fog, cloud and hybrid detectors detected attacks equally well within a few hundredths of F1. They differed in cost: the cloud in backhaul traffic, batching latency and sensitivity to backhaul loss, the fog in false alarms, which the hybrid's longer context reduced about 9-fold. What limited detection was observability. Flooding was almost always caught; spoofing by an attacker who suppresses the victim, FDI on noisy sensors and replay mostly were not, because they leave no evidence at any tier. Those attacks call for authentication, redundant sensing and physical watermarking rather than for moving the detector. Next steps are adaptive attackers, a radio-level link model, transfer across city topologies, and an implementation of watermark-based replay detection in the simulator.

\section*{Code and Data Availability}
\sloppy 
The simulator, experiment scripts, result files of every pipeline version (v0, v1 and the confirmatory v2), and the scripts that generate every table, figure and number in this paper are released under the MIT license at \url{https://github.com/mustafasadiq0/PY-CPS-IDS}. \texttt{experiments/simulate\_all.py}, \texttt{experiments/evaluate.py}, \texttt{experiments/breakdown.py} and \texttt{experiments/stats.py} reproduce all results (about 45 minutes on two CPU cores); \texttt{paper/make\_assets.py} and \texttt{paper/make\_text.py} regenerate the tables, figures and number-bearing text. An earlier version of this work, with a different simulator and without the confirmatory protocol, was submitted to a journal in 2026 and posted as a preprint on ResearchGate \citep{aljumaily2026pycpsrg}; this paper supersedes it.

\section*{Use of AI Tools}
An AI assistant (Claude, Anthropic) was used under the authors' direction to write and revise the simulator and experiment code, to run the experiments reported here, and to draft and edit the manuscript, including an internal review of the draft. The authors specified the research question, the design and the threat model, reviewed the code and the text, re-ran the released pipeline, and checked every reference against its source. The authors take full responsibility for the content.

\bibliographystyle{unsrtnat}
\bibliography{refs}

\begin{thebibliography}{34}
\providecommand{\natexlab}[1]{#1}
\providecommand{\url}[1]{\texttt{#1}}
\expandafter\ifx\csname urlstyle\endcsname\relax
  \providecommand{\doi}[1]{doi: #1}\else
  \providecommand{\doi}{doi: \begingroup \urlstyle{rm}\Url}\fi

\bibitem[Ghena et~al.(2014)Ghena, Beyer, Hillaker, Pevarnek, and
  Halderman]{ghena2014green}
Branden Ghena, William Beyer, Allen Hillaker, Jonathan Pevarnek, and J.~Alex
  Halderman.
\newblock Green lights forever: Analyzing the security of traffic
  infrastructure.
\newblock In \emph{Proc. 8th USENIX Workshop on Offensive Technologies (WOOT)},
  2014.

\bibitem[Ronen and Shamir(2016)]{ronen2016extended}
Eyal Ronen and Adi Shamir.
\newblock Extended functionality attacks on {IoT} devices: The case of smart
  lights.
\newblock In \emph{Proc. IEEE European Symposium on Security and Privacy
  (EuroS\&P)}, 2016.

\bibitem[Zarpel{\~a}o et~al.(2017)Zarpel{\~a}o, Miani, Kawakani, and
  de~Alvarenga]{zarpelao2017survey}
Bruno~Bogaz Zarpel{\~a}o, Rodrigo~Sanches Miani, Cl{\'a}udio~Toshio Kawakani,
  and Sean~Carlisto de~Alvarenga.
\newblock A survey of intrusion detection in {Internet of Things}.
\newblock \emph{Journal of Network and Computer Applications}, 84:\penalty0
  25--37, 2017.

\bibitem[Ferrag et~al.(2022)Ferrag, Friha, Hamouda, Maglaras, and
  Janicke]{ferrag2022edge}
Mohamed~Amine Ferrag, Othmane Friha, Djallel Hamouda, Leandros Maglaras, and
  Helge Janicke.
\newblock {Edge-IIoTset}: A new comprehensive realistic cyber security dataset
  of {IoT} and {IIoT} applications for centralized and federated learning.
\newblock \emph{IEEE Access}, 10:\penalty0 40281--40306, 2022.
\newblock \doi{10.1109/ACCESS.2022.3165809}.

\bibitem[Koroniotis et~al.(2019)Koroniotis, Moustafa, Sitnikova, and
  Turnbull]{koroniotis2019botiot}
Nickolaos Koroniotis, Nour Moustafa, Elena Sitnikova, and Benjamin Turnbull.
\newblock Towards the development of realistic botnet dataset in the {Internet
  of Things} for network forensic analytics: {Bot-IoT} dataset.
\newblock \emph{Future Generation Computer Systems}, 100:\penalty0 779--796,
  2019.

\bibitem[Alsaedi et~al.(2020)Alsaedi, Moustafa, Tari, Mahmood, and
  Anwar]{alsaedi2020toniot}
Abdullah Alsaedi, Nour Moustafa, Zahir Tari, Abdun Mahmood, and Adnan Anwar.
\newblock {TON\_IoT} telemetry dataset: A new generation dataset of {IoT} and
  {IIoT} for data-driven intrusion detection systems.
\newblock \emph{IEEE Access}, 8:\penalty0 165130--165150, 2020.
\newblock \doi{10.1109/ACCESS.2020.3022862}.

\bibitem[Raza et~al.(2013)Raza, Wallgren, and Voigt]{raza2013svelte}
Shahid Raza, Linus Wallgren, and Thiemo Voigt.
\newblock {SVELTE}: Real-time intrusion detection in the {Internet of Things}.
\newblock \emph{Ad Hoc Networks}, 11\penalty0 (8):\penalty0 2661--2674, 2013.
\newblock \doi{10.1016/j.adhoc.2013.04.014}.

\bibitem[Bonomi et~al.(2012)Bonomi, Milito, Zhu, and Addepalli]{bonomi2012fog}
Flavio Bonomi, Rodolfo Milito, Jiang Zhu, and Sateesh Addepalli.
\newblock Fog computing and its role in the internet of things.
\newblock In \emph{Proc. 1st Edition of the MCC Workshop on Mobile Cloud
  Computing}, pages 13--16, 2012.
\newblock \doi{10.1145/2342509.2342513}.

\bibitem[Diro and Chilamkurti(2018)]{diro2018distributed}
Abebe~Abeshu Diro and Naveen Chilamkurti.
\newblock Distributed attack detection scheme using deep learning approach for
  {Internet of Things}.
\newblock \emph{Future Generation Computer Systems}, 82:\penalty0 761--768,
  2018.
\newblock \doi{10.1016/j.future.2017.08.043}.

\bibitem[Mirsky et~al.(2018)Mirsky, Doitshman, Elovici, and
  Shabtai]{mirsky2018kitsune}
Yisroel Mirsky, Tomer Doitshman, Yuval Elovici, and Asaf Shabtai.
\newblock Kitsune: An ensemble of autoencoders for online network intrusion
  detection.
\newblock In \emph{Proc. Network and Distributed System Security Symposium
  (NDSS)}, 2018.
\newblock \doi{10.14722/ndss.2018.23204}.

\bibitem[Meidan et~al.(2018)Meidan, Bohadana, Mathov, Mirsky, Shabtai,
  Breitenbacher, and Elovici]{meidan2018nbaiot}
Yair Meidan, Michael Bohadana, Yael Mathov, Yisroel Mirsky, Asaf Shabtai,
  Dominik Breitenbacher, and Yuval Elovici.
\newblock {N-BaIoT}---network-based detection of {IoT} botnet attacks using
  deep autoencoders.
\newblock \emph{IEEE Pervasive Computing}, 17\penalty0 (3):\penalty0 12--22,
  2018.
\newblock \doi{10.1109/MPRV.2018.03367731}.

\bibitem[Antonakakis et~al.(2017)Antonakakis, April, Bailey, Bernhard,
  Bursztein, Cochran, Durumeric, Halderman, Invernizzi, Kallitsis, Kumar,
  Lever, Ma, Mason, Menscher, Seaman, Sullivan, Thomas, and
  Zhou]{antonakakis2017mirai}
Manos Antonakakis, Tim April, Michael Bailey, Matt Bernhard, Elie Bursztein,
  Jaime Cochran, Zakir Durumeric, J.~Alex Halderman, Luca Invernizzi, Michalis
  Kallitsis, Deepak Kumar, Chaz Lever, Zane Ma, Joshua Mason, Damian Menscher,
  Chad Seaman, Nick Sullivan, Kurt Thomas, and Yi~Zhou.
\newblock Understanding the {Mirai} botnet.
\newblock In \emph{Proc. 26th USENIX Security Symposium}, pages 1093--1110,
  2017.

\bibitem[Chen et~al.(2010)Chen, Yang, Trappe, and Martin]{chen2010identity}
Yingying Chen, Jie Yang, Wade Trappe, and Richard~P. Martin.
\newblock Detecting and localizing identity-based attacks in wireless and
  sensor networks.
\newblock \emph{IEEE Transactions on Vehicular Technology}, 59\penalty0
  (5):\penalty0 2418--2434, 2010.

\bibitem[Aljumaily et~al.(2025)Aljumaily, Abdullah, and
  Abd~Alhasan]{aljumaily2025hybrid}
Mustafa~S. Aljumaily, Sherwan~J. Abdullah, and Ahmed~Q. Abd~Alhasan.
\newblock A hybrid intrusion detection framework for cyber-physical security in
  smart home/smart city {IoT} systems.
\newblock \emph{International Journal of Mechatronics, Robotics, and Artificial
  Intelligence}, 1\penalty0 (2):\penalty0 63--73, 2025.
\newblock \doi{10.33971/ijmrai.1.2.9}.

\bibitem[Liu et~al.(2009)Liu, Ning, and Reiter]{liu2009fdi}
Yao Liu, Peng Ning, and Michael~K. Reiter.
\newblock False data injection attacks against state estimation in electric
  power grids.
\newblock In \emph{Proc. ACM Conf. Computer and Communications Security (CCS)},
  pages 21--32, 2009.

\bibitem[Mo and Sinopoli(2009)]{mo2009replay}
Yilin Mo and Bruno Sinopoli.
\newblock Secure control against replay attacks.
\newblock In \emph{Proc. 47th Annual Allerton Conf. Communication, Control, and
  Computing}, pages 911--918, 2009.
\newblock \doi{10.1109/ALLERTON.2009.5394956}.

\bibitem[Giraldo et~al.(2018)Giraldo, Urbina, Cardenas, Valente, Faisal, Ruths,
  Tippenhauer, Sandberg, and Candell]{giraldo2018survey}
Jairo Giraldo, David Urbina, Alvaro Cardenas, Junia Valente, Mustafa Faisal,
  Justin Ruths, Nils~Ole Tippenhauer, Henrik Sandberg, and Richard Candell.
\newblock A survey of physics-based attack detection in cyber-physical systems.
\newblock \emph{ACM Computing Surveys}, 51\penalty0 (4):\penalty0 76:1--76:36,
  2018.
\newblock \doi{10.1145/3203245}.

\bibitem[Urbina et~al.(2016)Urbina, Giraldo, Cardenas, Tippenhauer, Valente,
  Faisal, Ruths, Candell, and Sandberg]{urbina2016stealthy}
David~I. Urbina, Jairo~A. Giraldo, Alvaro~A. Cardenas, Nils~Ole Tippenhauer,
  Junia Valente, Mustafa Faisal, Justin Ruths, Richard Candell, and Henrik
  Sandberg.
\newblock Limiting the impact of stealthy attacks on industrial control
  systems.
\newblock In \emph{Proc. ACM SIGSAC Conf. Computer and Communications Security
  (CCS)}, pages 1092--1105, 2016.
\newblock \doi{10.1145/2976749.2978388}.

\bibitem[Mo et~al.(2015)Mo, Weerakkody, and Sinopoli]{mo2015watermark}
Yilin Mo, Sean Weerakkody, and Bruno Sinopoli.
\newblock Physical authentication of control systems: Designing watermarked
  control inputs to detect counterfeit sensor outputs.
\newblock \emph{IEEE Control Systems Magazine}, 35\penalty0 (1):\penalty0
  93--109, 2015.
\newblock \doi{10.1109/MCS.2014.2364724}.

\bibitem[Satchidanandan and Kumar(2017)]{satchidanandan2017dynamic}
Bharadwaj Satchidanandan and P.~R. Kumar.
\newblock Dynamic watermarking: Active defense of networked cyber--physical
  systems.
\newblock \emph{Proceedings of the IEEE}, 105\penalty0 (2), 2017.
\newblock \doi{10.1109/JPROC.2016.2575064}.

\bibitem[Sommer and Paxson(2010)]{sommer2010outside}
Robin Sommer and Vern Paxson.
\newblock Outside the closed world: On using machine learning for network
  intrusion detection.
\newblock In \emph{Proc. IEEE Symposium on Security and Privacy}, pages
  305--316, 2010.
\newblock \doi{10.1109/SP.2010.25}.

\bibitem[Arp et~al.(2022)Arp, Quiring, Pendlebury, Warnecke, Pierazzi,
  Wressnegger, Cavallaro, and Rieck]{arp2022dos}
Daniel Arp, Erwin Quiring, Feargus Pendlebury, Alexander Warnecke, Fabio
  Pierazzi, Christian Wressnegger, Lorenzo Cavallaro, and Konrad Rieck.
\newblock Dos and don'ts of machine learning in computer security.
\newblock In \emph{Proc. 31st USENIX Security Symposium}, 2022.

\bibitem[Riley and Henderson(2010)]{riley2010ns3}
George~F. Riley and Thomas~R. Henderson.
\newblock The ns-3 network simulator.
\newblock In Klaus Wehrle, Mesut G{\"u}ne{\c{s}}, and James Gross, editors,
  \emph{Modeling and Tools for Network Simulation}, pages 15--34. Springer,
  2010.
\newblock \doi{10.1007/978-3-642-12331-3_2}.

\bibitem[Varga and Hornig(2008)]{varga2008omnet}
Andr{\'a}s Varga and Rudolf Hornig.
\newblock An overview of the {OMNeT++} simulation environment.
\newblock In \emph{Proc. 1st Int. Conf. Simulation Tools and Techniques for
  Communications, Networks and Systems (SIMUTools)}, Marseille, France, 2008.
\newblock \doi{10.4108/ICST.SIMUTOOLS2008.3027}.

\bibitem[{\"O}sterlind et~al.(2006){\"O}sterlind, Dunkels, Eriksson, Finne, and
  Voigt]{osterlind2006cooja}
Fredrik {\"O}sterlind, Adam Dunkels, Joakim Eriksson, Niclas Finne, and Thiemo
  Voigt.
\newblock Cross-level sensor network simulation with {COOJA}.
\newblock In \emph{Proc. 31st IEEE Conference on Local Computer Networks
  (LCN)}, pages 641--648, 2006.
\newblock \doi{10.1109/LCN.2006.322172}.

\bibitem[Gupta et~al.(2017)Gupta, Dastjerdi, Ghosh, and
  Buyya]{gupta2017ifogsim}
Harshit Gupta, Amir~Vahid Dastjerdi, Soumya~K. Ghosh, and Rajkumar Buyya.
\newblock {iFogSim}: A toolkit for modeling and simulation of resource
  management techniques in the {Internet of Things}, edge and fog computing
  environments.
\newblock \emph{Software: Practice and Experience}, 47\penalty0 (9):\penalty0
  1275--1296, 2017.

\bibitem[Lopez et~al.(2018)Lopez, Behrisch, Bieker-Walz, Erdmann,
  Fl{\"o}tter{\"o}d, Hilbrich, L{\"u}cken, Rummel, Wagner, and
  Wie{\ss}ner]{lopez2018sumo}
Pablo~Alvarez Lopez, Michael Behrisch, Laura Bieker-Walz, Jakob Erdmann,
  Yun-Pang Fl{\"o}tter{\"o}d, Robert Hilbrich, Leonhard L{\"u}cken, Johannes
  Rummel, Peter Wagner, and Evamarie Wie{\ss}ner.
\newblock Microscopic traffic simulation using {SUMO}.
\newblock In \emph{Proc. 21st IEEE Int. Conf. Intelligent Transportation
  Systems (ITSC)}, pages 2575--2582, 2018.
\newblock \doi{10.1109/ITSC.2018.8569938}.

\bibitem[Sch{\"u}tte et~al.(2011)Sch{\"u}tte, Scherfke, and
  Tr{\"o}schel]{schutte2011mosaik}
Steffen Sch{\"u}tte, Stefan Scherfke, and Martin Tr{\"o}schel.
\newblock Mosaik: A framework for modular simulation of active components in
  smart grids.
\newblock In \emph{Proc. IEEE First Int. Workshop on Smart Grid Modeling and
  Simulation (SGMS)}, pages 55--60, 2011.

\bibitem[Mathur and Tippenhauer(2016)]{mathur2016swat}
Aditya~P. Mathur and Nils~Ole Tippenhauer.
\newblock {SWaT}: A water treatment testbed for research and training on {ICS}
  security.
\newblock In \emph{Proc. Int. Workshop on Cyber-physical Systems for Smart
  Water Networks (CySWater)}, pages 31--36, 2016.

\bibitem[{SimPy Developers}(n.d.)]{simpy}
{SimPy Developers}.
\newblock {SimPy}: Discrete event simulation for {Python}, version 4.1.
\newblock \url{https://simpy.readthedocs.io/}, n.d.
\newblock Accessed 26 Nov. 2026.

\bibitem[Karanjkar et~al.(2019)Karanjkar, Tejasvi, and
  Amrutur]{karanjkar2019simpy}
Neha Karanjkar, Poorna~Chandra Tejasvi, and Bharadwaj Amrutur.
\newblock A {SimPy}-based simulation testbed for smart-city {IoT} applications:
  Poster abstract.
\newblock In \emph{Proc. 4th ACM/IEEE Int. Conf. Internet of Things Design and
  Implementation (IoTDI)}, pages 273--274, 2019.

\bibitem[Breiman(2001)]{breiman2001rf}
Leo Breiman.
\newblock Random forests.
\newblock \emph{Machine Learning}, 45\penalty0 (1):\penalty0 5--32, 2001.
\newblock \doi{10.1023/A:1010933404324}.

\bibitem[Pedregosa et~al.(2011)Pedregosa, Varoquaux, Gramfort, Michel, Thirion,
  Grisel, Blondel, Prettenhofer, Weiss, Dubourg, Vanderplas, Passos,
  Cournapeau, Brucher, Perrot, and Duchesnay]{pedregosa2011sklearn}
Fabian Pedregosa, Ga{\"e}l Varoquaux, Alexandre Gramfort, Vincent Michel,
  Bertrand Thirion, Olivier Grisel, Mathieu Blondel, Peter Prettenhofer, Ron
  Weiss, Vincent Dubourg, Jake Vanderplas, Alexandre Passos, David Cournapeau,
  Matthieu Brucher, Matthieu Perrot, and {\'E}douard Duchesnay.
\newblock Scikit-learn: Machine learning in {Python}.
\newblock \emph{Journal of Machine Learning Research}, 12:\penalty0 2825--2830,
  2011.

\bibitem[Aljumaily and Abd(2026)]{aljumaily2026pycpsrg}
Mustafa~S. Aljumaily and Hayder~Kareem Abd.
\newblock {PyCPS-IDS}: A {Python}-based cyber-physical simulation framework for
  {AI}-driven intrusion detection in smart city {IoT} networks.
\newblock Preprint, ResearchGate,
  \href{https://doi.org/10.13140/RG.2.2.19297.77923}{doi:10.13140/RG.2.2.19297.77923},
  2026.

\end{thebibliography}
\end{document}